# Real-space observation of short-period cubic lattice of skyrmions in MnGe

Toshiaki Tanigaki,[1,2,*] Kiyou Shibata,[3] Naoya Kanazawa,[3] Xiuzhen Yu,[1] Sinji Aizawa,[1] Yoshinori Onose,[4] Hyun Soon Park,[1,†] Daisuke Shindo,[1,5] and Yoshinori Tokura.[1,3]

[1] RIKEN Center for Emergent Matter Science (CEMS), Wako 351-0198, Japan

[2] Central Research Laboratory, Hitachi, Ltd., Hatoyama 350-0395, Japan

[3] Department of Applied Physics, University of Tokyo, Tokyo 113-8656, Japan

[4] Department of Basic Science, University of Tokyo, Tokyo 153-8902, Japan.

[5] Institute of Multidisciplinary Research for Advanced Materials, Tohoku University, Sendai 980-8577, Japan

[†]Present address: Department of Materials Science & Engineering, Dong-A University, Busan 604-714, Republic of Korea

*e-mail: toshiaki.tanigaki.mv@hitachi.com

**Emergent phenomena and functions arising from topological electron-spin textures in real space or momentum space are attracting growing interest for new concept of states of matter as well as for possible applications to spintronics[1-5]. One such example is a magnetic skyrmion[3-5], a topologically stable nanoscale spin vortex structure characterized by a topological index. Real-space regular arrays of skyrmions are described by combination of multi-directional spin helixes. Nanoscale configurations and characteristics of the two-dimensional skyrmion hexagonal-lattice have been revealed extensively by real-space observations[6-8]. Other three-dimensional forms of skyrmion lattices, such as a cubic-lattice of skyrmions, are also anticipated to exist[9,10], yet their direct observations remain elusive. Here we report real-space observations of spin configurations of the skyrmion cubic-lattice in MnGe with a very short period (~3 nm) and hence endowed with the largest skyrmion number density. The skyrmion lattices parallel to the {100} atomic lattices are directly observed using Lorentz transmission electron microscopes (Lorentz TEMs). It enables the first simultaneous observation of magnetic skyrmions and underlying atomic-lattice fringes. These results indicate the emergence of skyrmion-antiskyrmion lattice in MnGe, which is a source of emergent electromagnetic responses[9,11] and will open a possibility of controlling few-nanometer scale skyrmion lattices through atomic lattice modulations.**

The discovery of giant magnetoresistance[12,13] in magnetic multilayers opened a new field, i.e., spintronics, of controlling motions of electrons using structures of spin ensemble[14]. In solids, spins interact with electron orbits at specific atomic sites. This spin-orbit (or spin-orbital) interaction is the origin of emergent states and phenomena, such as colossal magnetoresistance[15], topological insulators[1,2], spin glasses[16], and magnetic skyrmions[3-5]. Among these, magnetic skyrmions found in helimagnets, for example those having B20 type (MnSi type) structure, are expected to be applied as information carriers in low-power-consumption spintronics memories[17,18] by taking advantages of their pinning-free motions with low excitation current densities (~$10^6$ A/m$^2$)[19,20]. Skyrmions are topologically characterized as composed of swirling spins which point to all directions wrapping a sphere integer times. This winding number (integer, usually unity) is called the skyrmion number and plays an essential role in the magneto-charge-transport as well as in the skyrmion dynamics itself. An emergent magnetic field created by the skyrmion number induces a nontrivial Hall effect termed topological Hall effect (THE)[21,22], while the gyrocoupling term in the Thiele equation[23,24] as represented by the skyrmion number enables the nearly pinning-free motion of skyrmion driven by spin current or spin-polarized charge current in magnets. Since the emergent magnetic field is proportional to the skyrmion density, higher-density skyrmions are considered to show a larger THE.

Spin helical structures, including linear helixes and skyrmions, in chiral-lattice magnets are caused by the competition between symmetric (ferromagnetic) exchange interaction *J* and antisymmetric exchange (Dzyaloshinskii-Moriya) interaction *D*, the latter of which arises from the relativistic spin-orbit coupling in the non-centrosymmetric crystal structure, *e.g.* of B20 compounds (space group $P2_13$)[25,26]. The helical period $L$ is proportional to $J/D$. Small-angle neutron scattering (SANS) studies showed linear (i.e. single wave vector **Q**) helix structures (Fig. 1a) in B20 type helimagnets; bulk MnSi ($L$ = 18 nm) [27], $Fe_{1-x}Co_xSi$ ($L$ = 43-230 nm)[28], $Co_{1-x}Mn_xSi$ ($L$ = 11-14 nm)[28], and FeGe ($L$ = 68-70 nm)[29]. For these materials, Lorentz transmission electron microscopy has enabled the real-space observation of local magnetic-moment configurations or helical spin textures, including bending and edge-dislocations of the linear helixes[30,31]. In Lorentz transmission electron microscope (TEM) observation, magnetic-moments with opposite directions in the helical configurations deflect transmitting electrons in the opposite directions, generating contrasts due to magnetic origin (Fig. 1e)[31]. Note that those contrasts reverse when focusing conditions are reversed from *underfocus* to *overfocus*. In the chiral-lattice helimagnets, the spin swirling direction (magnetic helicity) is determined by underlying crystal chirality[32].

Two-dimensional (2D) lattices of skyrmions are composed of coplanar multiple wave vectors of helical magnetic-moment configurations[4,5]. The 2D hexagonal lattices of skyrmions (Fig. 1b) were detected by SANS in several B20 type compounds[4,33,34]. They are composed of triangular wave vectors (triple-**Q**) formed in the plane perpendicular to the external magnetic field *H*. Real-space observations using Lorentz TEM have clearly shown not only local magnetic-moment configurations of skyrmions but also the thickness sensitivity of the skyrmion phase diagram[6-8]. In the 2D skyrmion lattice category, square lattice composed of rectangular wave vectors (double-**Q**; Fig. 1c) have been proposed on the basis of simulations[3,35-37], yet their existence has not been confirmed experimentally.

Recently, a short-period magnetic-moment configuration ($L$ = 3-6 nm) with wave vectors perpendicular to the {100} plane has been found in MnGe by a powder neutron diffraction mehtod[9]. MnGe shows the largest THE among the B20 compounds (40 times as large as that of MnSi) over the whole helimagnetic phase in the *T-H* plane ($0 < T <$ 170 K and $0 < H <$ 12 T). Lowering of crystal symmetry from cubic to orthorhombic is observed upon the magnetic transition by powder neutron diffraction studies[38]. The results of SANS and topological Hall resistivity measurements imply that the existence of a three-dimensional (3D) cubic lattice of skyrmions (Fig. 1d) coupled to the underlying atomic crystal lattice in MnGe[10].

The 3D lattice of skyrmions is anticipated to be composed of non-coplanar multiple wave vectors **Q**$_i$ of helical magnetic-moment configurations. The configurations of magnetic moments **m** for the most simple case can be expressed by[5,36]

$$\mathbf{m}(\mathbf{r}) = \mathbf{m_0} + \sum_{i=1}^{3} \mathbf{m}_{\mathbf{Q}_i}(\mathbf{r} + \Delta \mathbf{r}_i) , \qquad (1)$$

where $\mathbf{r}$ is the three-dimensional position vector, and $\mathbf{m_0}$ the uniform magnetization, $\mathbf{m}_{\mathbf{Q}_i}(\mathbf{r}) = A_i[\mathbf{m}_{i1}\cos(\mathbf{Q}_i \cdot \mathbf{r}) + \mathbf{m}_{i2}\sin(\mathbf{Q}_i \cdot \mathbf{r})]$. Here, $A_i$ and $\mathbf{Q}_i \cdot \Delta \mathbf{r}_i$ are the amplitude and phase of each helix, respectively; $\mathbf{m}_{i1}$ and $\mathbf{m}_{i2}$ are perpendicular to each other and $\mathbf{Q}_i$. In particular, the cubic-lattice skyrmions is composed of three orthogonal wave vectors $\mathbf{Q}_i$. Since the wave vectors $\mathbf{Q}_i$ in MnGe is perpendicular to the {100} plane, two-directional Lorentz TEM contrasts (Fig. 1h) are expected to be observed in the [001] oriented MnGe plate.

The real-space observation of cubic-lattice skyrmions is necessary to understand the local magnetic-moment configurations and to develop spintronics devices utilizing this type of robust skyrmions carrying a large emergent magnetic flux. However, real-space observations and analyses on the magnetic-moment configurations and their relation to underlying atomic crystal lattices in MnGe have not yet been performed because of difficulty in single-crystal synthesis[9,10,38,39] and sub-nm-resolution Lorentz TEM observations. Here we report the first real-space observations of the short-period (~ 3 nm) cubic-lattice skyrmions in MnGe using Lorentz TEM. The [001]-oriented thin samples were prepared from a polycrystalline MnGe sample using a microsampling method and observed with magnetic fields $H$ of 0 T and 2.4 T applied perpendicular to the thin sample, using two types of TEMs (for the detail, see Methods section and Supplemental information). For both observation conditions, the cubic-lattice skyrmions are observed and the temperature dependence of the period $L$ (or skyrmion lattice constant) is obtained and compared with the neutron diffraction results. We note that, the magnetic skyrmions and crystal lattice fringes in MnGe can be simultaneously imaged using high-resolution Lorentz TEM, as shown below.

Figure 2 shows the Lorentz TEM images of the [001]-oriented MnGe thin sample at $H = 0$ and 2.4 T applied perpendicular to the sample plane. Crystal orientations were determined from electron diffraction patterns in magnetic field-free Lorentz TEM results and crystal lattice fringes obtained at 2.4 T in high-resolution Lorentz TEM observation. The orthogonal distortion of the crystal lattice reported in the powder neutron diffraction analysis[38] was not detected within the accuracy of the present TEM observation; therefore the lattices are indexed in the cubic setting in this report. Two-directional stripe contrasts along the [100] and [010] directions of underlying crystal lattice were observed at 20 K (Fig. 2a and d) and 70 K (Fig. 2b and e), while they disappeared above $T_N \approx 170$ K for 0 T (Fig. 2c) and at 120 K for 2.4 T (Fig. 2f); the observed temperature dependence at the respective magnetic fields ensures that the observed contrasts are of magnetic origin. The crystalline state of skyrmions in MnGe appears to exist in a very similar form at both zero and applied magnetic fields, as far as the samples are located in the phase diagram region of temperature and magnetic-field for the helimagnetic phase[9]. The observed cross stripe image results are in good agreement with those expected in the Lorentz TEM observation for the cubic-lattice skyrmions (Fig. 1h), while these results are also similar to those of the 2D square-lattice (Fig. 1g). Nevertheless, a 2D skyrmion square-lattice in MnGe is unlikely to happen; the results of SANS showed clearly the presence of the wave vector parallel to the applied magnetic field dirction[10], which corresponds to the wave vector normal to the plane in the present configuration under the magnetic field of 2.4 T. These observed cross stripes in MnGe are always straight without any edge dislocations or bendings, in contrast to those of curved stripes observed for the helical state in $Fe_{0.5}Co_{0.5}Si$[6,30] and FeGe[7,31]. Skyrmion lattices in MnGe are always parallel to the {100} lattice planes, while a variety helix stripes and/or skyrmion lattice configurations appear in MnSi[40]. These observed results may reflect the fact that the magnetic anisotropy of MnGe is larger than that of other B20-type skyrmion-hosting compounds, *e.g.,* $Fe_{0.5}Co_{0.5}Si$, FeGe, and MnSi. Such strong magnetic anisotropy may come from the short lattice constants of skyrmions, only several times the atomic lattice constant.

Figure 2h shows the temperature dependence of the magnetic-moment configuration period (skyrmion lattice priod) $L$ for the perpendicular magnetic fields (0 T and 2.4 T). The period remains constant at around 3 nm below 80 K under the both experimental conditions, while above 80 K the periods gradually increase with temperature. These results are in good agreement with those obtained by powder neutron diffraction and SANS experiments[9,10,38]. In consideration of the temperature($T$)-independent relation expected for a simple chiral-lattice helimagnet[25,26], $L \sim aJ/D$ ($a$: the lattice constant), the magnetic anisotropy must play an additional role in this anomalous $T$-dependence of $L$ up to $T_N$.

Figure 3a exemplifies a success in getting the high-resolution Lorentz TEM image both for magnetic-moment configurations and atomic crystal lattice, obtained at 35K and 2.4T in MnGe. The cross stripe contrasts with a period of 3 nm of cubic-lattice skyrmions and the atomic lattice fringes of the (100) and (010) planes were simultaneously observed in the underfocus condition with a defocus of 3.0±0.3 μm. Even though atomic positions cannot be determined precisely from this image due to defocus conditions and a spherical aberration of the lens of the electron microscope, the magnetic-moment ordering periods and directions relative to the atomic crystal lattices can be clearly indicated. Here it is worth noting the difference between observed intensities for respective **Q** vectors. Right and left directional magnetic contrast along the (010) plane were slightly weaker than that along the (100) plane, as seen in Fig. 3a and also noticed in the area surrounded by the white rectangle in Fig. 2d. We conjecture the following possibilities as the cause for the unequal Lorentz TEM contrasts: sample tilting from the [001] orientation, a tilting of the **Q** vector from the crystal axis, and unequal amplitude for each magnetic-moment helix. In the experiment, the sample tilting was adjusted in terms of the observed electron diffraction patterns and hence the **Q** vector tilting from the crystal axis should be minimal, less than a few degrees. In the thin film observation (thickness ~30 nm) within such a small tilt angle, the tilting effects, i.e., overlapping projections, are small for the 3 nm spaced structures. Thus, the local amplitude fluctuations for each Q vector are likely to exist as an intrinsic origin or a lattice-strain or an orthorhombicity dependent feature. A 2D Monte Carlo simulation by varying anisotropy strength and magnetic field strength showed that square-lattice skyrmions with unequal amplitude for each helix can be stabilized at large anisotropy rather than the hexagonal-lattice skyrmions[37]. The 3D lattice of skyrmions may show a similar trend, although this should be confirmed by large-scale 3D simulations.

Figure 3b shows in-plane magnetic-moment configurations obtained with the transport-of-intensity equation (TIE) in the QPt software package (HREM Co.)[41] using the Lorentz TEM image data obtained at 20 K under 2.4 T applied perpendicular to the sample with defocus of 2.5±0.3 μm (Fig. 2d). White arrows and color wheel indicate magnetic-moment directions and intensities projected onto the sample plane, respectively. Two-directional magnetic-moments along the [100] and [010] directions of underlying crystal lattice were clearly observed, although it is difficult to determine quantitative values of magnetic moments.

Figure 3c shows the enlarged image of the in-plane magnetic-moment texture (Fig. 2) to compare the simulated image based on the simplest cubic skyrmion crystal as represented by Eq. (1); the agreement is excellent, ensuring again the presence of the cubic lattice of skyrmions in MnGe. For 2D skyrmion, vortex-like (circling-magnetic-moment) areas correspond to the skyrmion positions. Note that, however, this is not for the 3D skyrmion cubic-lattice[36]; It contains both hedgehog and anti-hedgehog spin textures (see Figs. 3e and 3f), which carry the winding numbers of +1 and -1, respectively. The simulated in-plane magnetic-moment configurations results (Fig. 3d) also show the projected positions of hedgehog (blue) and anti-hedgehog (red). Note that those topological spin textures cannot be explicitly observed in the averaged in-plane moment configuration because of their three-dimensionally-modulated spin arrangement. Since the sign of the skyrmion number is opposite for hedgehog and anti-hedgehog, the total skyrmion number is cancelled out to be zero at the ground state for the zero external magnetic field. However, the application of magnetic field can induce the large skyrmion-number density due to the first term (Zeeman term) of Eq. (1), generating acorrespondingly large topological Hall effect as observed[9].

In conclusion, real-space observations using Lorentz TEM have revealed structural characteristics of cubic-lattice skyrmions in a helimagnet MnGe. The temperature dependence of the magnetic-moment configuration period $L$ shows the identical $T$-dependent behavior with the neutron scattering results. The simultaneous observation of atomic crystal lattices and magnetic-moment configurations usinghigh-resolution Lorentz TEMs has shown that skyrmion lattices are tightly fixed parallel to the {100} planes. The observed atomic-lattice-coupled magnetic-moment configuration indicates a possibility of controlling skyrmion lattices in terms of the crystal lattice strains. The present observations of short-period skyrmion-antiskyrmion (or spin hedgehog-antihedgehog) crystal lattices are the first step toward visualizing spins at the atomic resolution using electron microscopy. Since the spin-orbit interaction causes a lot of intriguing phenomena with potential of application to spintronics, real-space observations of sub-nanometer-scale topological spin textures will be an important subject in the forthcoming study.

**Methods**

Polycrystalline MnGe samples were synthesized with a cubic-anvil-type high-pressure apparatus. First, a mixture of Mn and Ge powders with an atomic ratio of 1:1 was arc-melted in an Ar atmosphere. Then, the MnGe alloy was placed in a cylindrical BN capsule and heat-treated for 1 h at 1073 K under 4 GPa[9]. Thin samples for a pre-TEM observation were prepared using an Ar ion-thinning method. It requires a large number of [001]-oriented thin samples to clearly observe helical magnetic-moment configuration with wave vector **Q** perpendicular to the {100} in MnGe using a high-resolution Lorentz TEM with a helium cooling holder. However, suitable samples for observation are difficult to prepare because crystal grains tend to tilt more than 20 degrees from the [001] orientation direction, while the tilt angle of the helium cooling holder is up to 15 degrees in the TEM. To overcome these difficulties we developed a following procedure (see Supplemental information for more details). First, the [001] oriented grains were selected from polycrystalline samples by an electron diffraction method using high-tilt angle TEM with wide gap (10 nm) objective lens, and then the selected [001]-oriented micro samples were picked out using a micro-manipulator in the focused ion beam (FIB) instruments (NB5000, Hitachi High-Technologies Co.)[42]. These micro samples were fixed onto a Mo support plate with the [001] direction normal to the plate and thinned down to 200 nm using FIB instrument. Finally, the samples were thinned to 10-50 nm by an Ar ion beam instrument (Gentle Mill Hi, Technoorg Linda Ltd. Co.) operated at 500 V. Since the [001]-oriented grains were larger than 100 μm, it was possible to obtain multiple experimental samples from a single polycrystalline sample. Magnetic field of 2.4 T was applied perpendicular to the sample plane by the TEM magnetic electron lens. Observations were performed using a 300-kV cold-field-emission TEM (HF-3300X, Hitachi High-Technologies Co.) with a high-resolution objective lens; zero-magnetic-field observations were performed using a 300-kV cold-field-emission TEM (HF-3300S, Hitachi High-Technologies Co.) by placing the sample at the Lorentz position having a negligible magnetic field (the same order as the geomagnetic field owing to the permalloy shielding). A double-tilt helium cooling holder (ULTDT, Gatan) was used for low temperature observations. A charge-coupled-device (CCD) camera with 2048 × 2048 pixels (ORIUS® SC200, Gatan) was used to obtain Lorentz TEM images.

**Acknowledgements**

The authors are grateful to the late Dr. A. Tonomura, Dr. T. Matsuda, Dr. Y. Murakami, Mr. K. Yanagisawa, and Dr. Y. A. Ono for valuable discussions and supports. This research was supported by a grant from the Japan Society for the Promotion of Science (JSPS) through the "Funding Program for World-Leading Innovative R&D on Science and Technology (FIRST Program)," initiated by the Council for Science and Technology and Innovation (CSTI) under the programs 'Development and Application of an Atomic-resolution Holography Electron Microscope' and 'Quantum Science on Strong Correlation'.

**Author contributions**

T.T., K.S., X.Y., N.K., S.A., H.S.P. and Y.T. designed the experiment. T.T., K.S., X.Y. and S.A. performed the microscopic experiment and analyzed the data. N. K. and Y. O. synthesized the bulk samples. T.T. and Y.T. wrote the paper. All authors discussed the results and commented on the manuscript.

**Additional information**

Reprint and permission information is available online at www.nature.com/reprints.

Correspondence and requests for materials should be addressed to T.T.

**Competing financial interests**

The authors declare no competing financial interests.

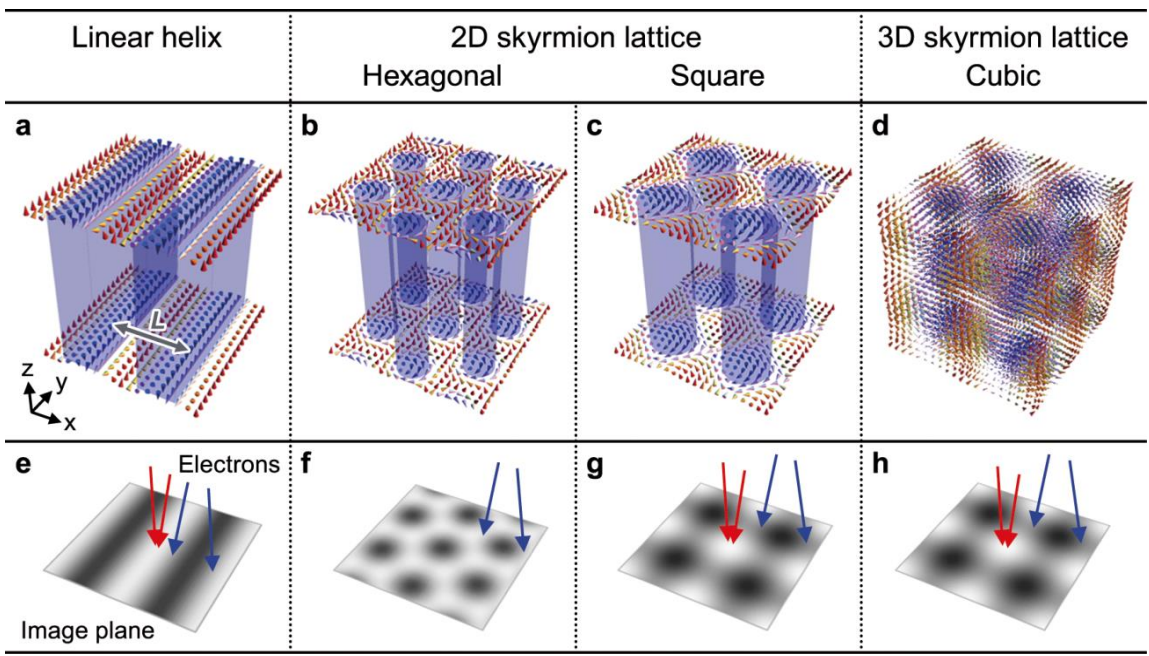

**Figure 1. Schematics of magnetic-moment configurations and simulated Lorentz transmission electron microscopy (Lorentz TEM) images. a-d**: 3D views of magnetic-moment configurations of a linear helix, two-dimensional (2D) hexagonal and square lattices of skyrmions, and a three-dimensional (3D) cubic lattice of three dimensional (3D) skyrmions. Arrows indicate magnetic moments. The hue of arrows represents the direction of z components; red (blue) arrows indicate up (down) magnetic moments. In the linear helix and 2D skyrmion lattices, magnetic-moment configurations are the same through the z direction. **e-h**: Anticipated overfocused Lorentz TEM images to be observed from the z direction. Transmitting electrons shown in red and blue arrows are deflected by Lorentz force due to helical magnetic-moment configurations and produce dark and light contrasts in the images. Blue plates, cylinders, and particle areas in linear helix, 2D and 3D skyrmion lattices correspond to magnetic-moment configurations producing dark contrasts in the images.

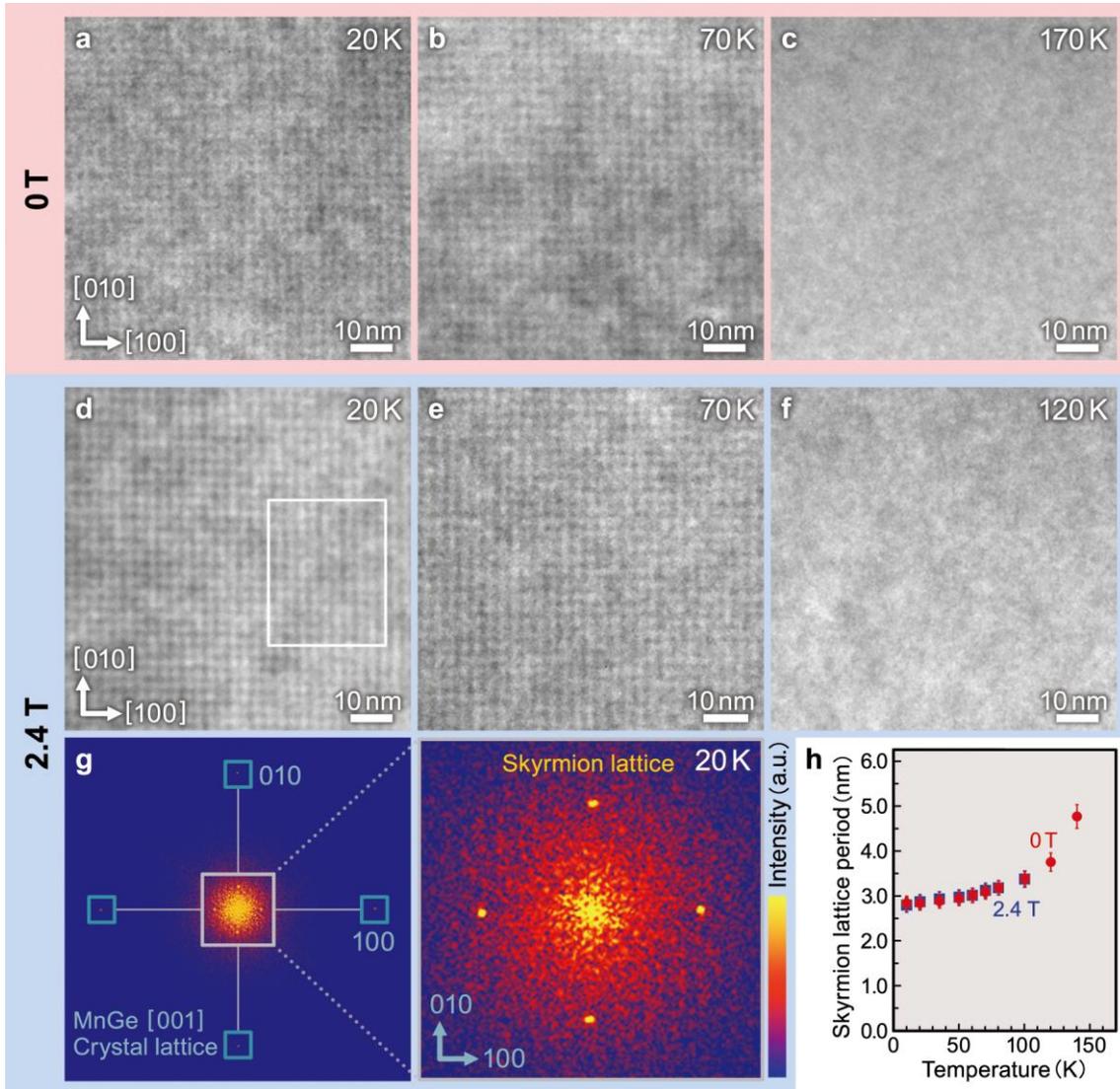

**Figure 2. Lorentz TEM images of cubic-lattice skyrmion in MnGe and temperature dependence of skyrmion lattice period.** Lorentz TEM images for the magnetic fields 0 T (**a-c**) and 2.4 T (**d-f**) applied perpendicular to the thin sample plane. (**a,d**) and (**b,e**): Two-directional stripe contrasts due to cubic-lattice skyrmion observed at 20 K and 70 K, respectively. Lorentz TEM contrasts disappear around $T_N$ = 170 K (0 T) and 120 K (2.4 T). Skyrmion lattices are parallel to the [100] and [010]. In some areas, Lorentz TEM contrasts along the [100] and [010] directions are different indicated by the white rectangle in **d**. **g**, Fast Fourier transformed image of **d** indicating parallel configurations of skyrmion lattices along the [100] and [010] directions. **h**, Temperature dependence of skyrmion lattice period for the magnetic field applied perpendicular to the thin sample plane (red circles, 0 T; and blue square, 2.4 T). The skyrmion lattice period increases with the increase in temperature.

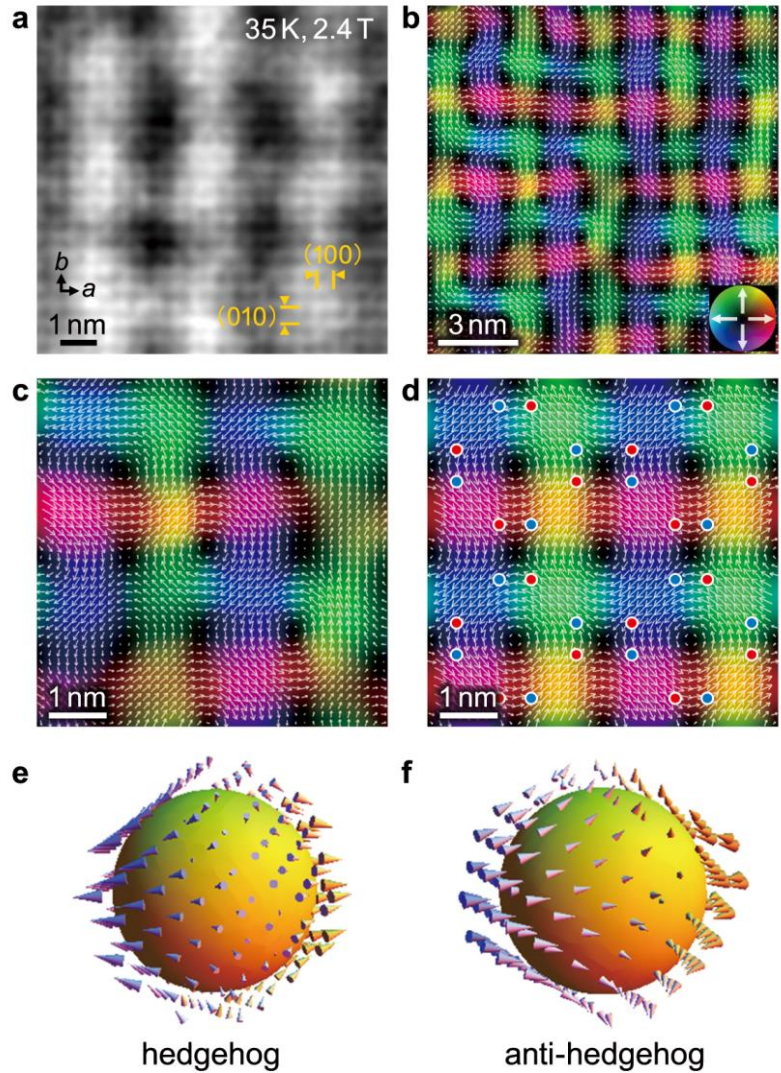

**Figure 3. In-plane magnetic-moment configurations of cubic-lattice skyrmion in MnGe and projected positions of hedgehog and anti-hedgehog. a**, Underfocused Lorentz TEM image of [001]-oriented MnGe at 35 K under 2.4 T applied perpendicular to the sample plane. Skyrmion lattices are simultaneously observed with the lattice fringes of the (100) and the (010) planes indicated by yellow lines. Arrows *a* and *b* indicate the *a* and *b* unit cells. **b**, Obtained magnetic-moment configurations at 20 K under 2.4 T applied perpendicular to the sample plane. White arrows and color wheel inserted at the bottom right indicate the directions of magnetic moments. **c**, Enlarged in-plane magnetic-moment configurations. **d**, Simulated in-plane magnetic-moment configurations and projected positions of hedgehogs (blue) and anti-hedgehogs (red). **e**, hedgehog spin texture. **f**, anti-hedgehog spin texture.

# Supplemental information

# Real-space observation of short-period cubic lattice of skyrmions in MnGe

Toshiaki Tanigaki, [1,2,*] Kiyou Shibata,[3] Naoya Kanazawa,[3] Xiuzhen Yu,[1] Sinji Aizawa,[1] Yoshinori Onose,[4] Hyun Soon Park,[1,†] Daisuke Shindo,[1,5] and Yoshinori Tokura.[1,3]

[1] RIKEN Center for Emergent Matter Science (CEMS), Wako 351-0198, Japan

[2] Central Research Laboratory, Hitachi, Ltd., Hatoyama 350-0395, Japan

[3] Department of Applied Physics, University of Tokyo, Tokyo 113-8656, Japan

[4] Department of Basic Science, University of Tokyo, Tokyo 153-8902, Japan.

[5] Institute of Multidisciplinary Research for Advanced Materials, Tohoku University, Sendai 980-8577, Japan

[†] Present address: Department of Materials Science & Engineering, Dong-A University, Busan 604-714, Republic of Korea

*e-mail: toshiaki.tanigaki.mv@hitachi.com

In this supplemental information, we describe the sample preparation process for observations of magnetic-moment configurations in MnGe. Lorentz transmission electron microscopic (TEM) observation of short-period helical magnetic-moment configurations using a helium cooling holder is difficult to perform. It requires a large number of [001]-oriented thin samples to clearly observe helical magnetic-moment configuration with wave vector **Q** perpendicular to the {100} in MnGe.. However, the thinned grains in polycrystalline tend to tilt their normal more than 20 degree from the [100] orientation. It is difficult to tilt them for proper orientation in the TEM, because the maximum tilt angle of the helium cooling holder is 15 degrees. To overcome these difficulties we developed the following method: First, thin samples for the pre-TEM observation were prepared using Ar ion-thinning method. Then, the [001] oriented grains were selected in polycrystalline samples by an electron diffraction method using a TEM with wide gap (10 nm) objective lens allowing 60 degree tilt angle (Figures S1a). Following this, the [001] oriented micro samples were picked out using a micro-manipulator in the focused ion beam (FIB) instruments (NB5000, Hitachi High-Technologies Co.) operated at 40 kV (Figures S1b)[1]. These micro samples were fixed onto a Mo support plate with the [001] direction normal to the plate using

W-deposition and thinned down to 200 nm using FIB instrument. Finally, to remove the surface damage layers induced by Ga ions of FIB, the samples were thinned to 10-50 nm by an Ar ion beam instrument (Gentle Mill Hi, Technoorg Linda Ltd. Co.) operated at 500 V (Figure S1c). Since the [001]-oriented grains were larger than 100 μm, it was possible to obtain plural thin samples from a single polycrystalline sample. Another benefit of this sample preparation technique is that the magnetic field can be applied exactly perpendicular to the thin sample plane in the selected crystal orientation.

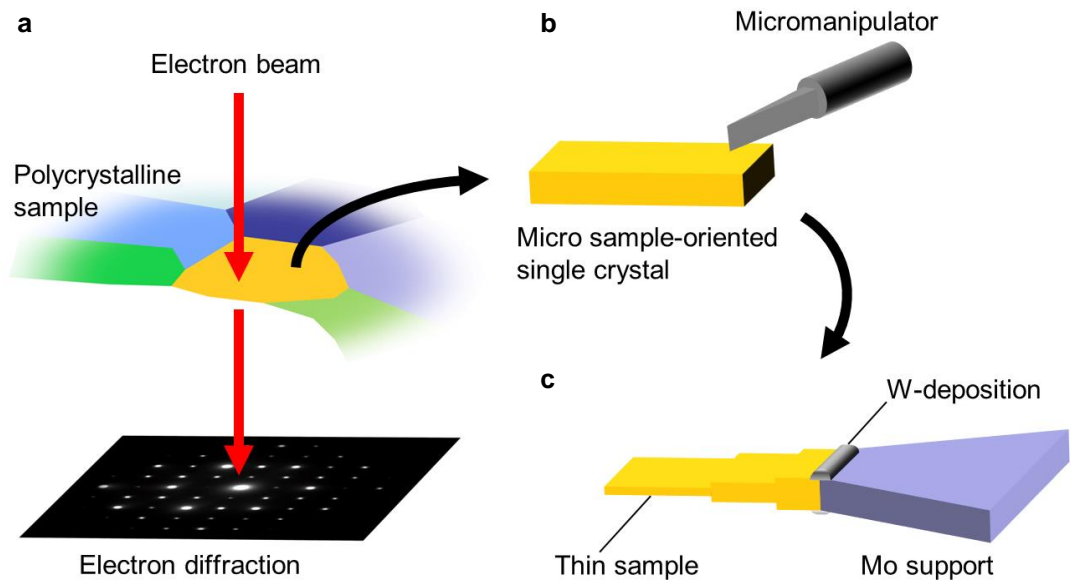

**Figure S1. Schematic presentations of orientation selective sampling.**
**a**, Transmission electron microscope (TEM) and electron diffraction observations for searching desired the [001] oriented grains in MnGe polycrystalline sample at room temperature. **b**, Micro samples were picked out from the selected crystal grain using micro-manipulator of focused ion beam instrument. **c**, The sample was fixed onto Mo support and thinned by using Ar ion beam for Lorentz TEM observations of skyrmion.